   \documentstyle[12pt]{article}
\if@twoside\oddsidemargin25mm
\else\oddsidemargin25mm\evensidemargin25mm\marginparwidth25mm\fi%
\begin{document}
\newcommand{\ft}[2]{{\textstyle\frac{#1}{#2}}}
\newcommand{\QED}{{\hspace*{\fill}\rule{2mm}{2mm}\linebreak}}
\def\dop{{\rm d}\hskip -1pt}
\def\bfone{\relax{\rm 1\kern-.35em 1}}
\def\bfzero{\relax{\rm I\kern-.18em 0}}
\def\inbar{\vrule height1.5ex width.4pt depth0pt}
\def\IC{\relax\,\hbox{$\inbar\kern-.3em{\rm C}$}}
\def\ID{\relax{\rm I\kern-.18em D}}
\def\IF{\relax{\rm I\kern-.18em F}}
\def\IK{\relax{\rm I\kern-.18em K}}
\def\IH{\relax{\rm I\kern-.18em H}}
\def\II{\relax{\rm I\kern-.17em I}}
\def\IN{\relax{\rm I\kern-.18em N}}
\def\IP{\relax{\rm I\kern-.18em P}}
\def\IQ{\relax\,\hbox{$\inbar\kern-.3em{\rm Q}$}}
\def\IR{\relax{\rm I\kern-.18em R}}
\def\IG{\relax\,\hbox{$\inbar\kern-.3em{\rm G}$}}
\font\cmss=cmss10 \font\cmsss=cmss10 at 7pt
\def\ZZ{\relax\ifmmode\mathchoice
{\hbox{\cmss Z\kern-.4em Z}}{\hbox{\cmss Z\kern-.4em Z}}
{\lower.9pt\hbox{\cmsss Z\kern-.4em Z}}
{\lower1.2pt\hbox{\cmsss Z\kern-.4em Z}}\else{\cmss Z\kern-.4em
Z}\fi}
\def\a{\alpha} \def\b{\beta} \def\d{\delta}
\def\e{\epsilon} \def\c{\gamma}
\def\G{\Gamma} \def\l{\lambda}
\def\L{\Lambda} \def\s{\sigma}
\def\cA{{\cal A}} \def\cB{{\cal B}}
\def\cC{{\cal C}} \def\cD{{\cal D}}
\def\cF{{\cal F}} \def\cG{{\cal G}}
\def\cH{{\cal H}} \def\cI{{\cal I}}
\def\cJ{{\cal J}} \def\cK{{\cal K}}
\def\cL{{\cal L}} \def\cM{{\cal M}}
\def\cN{{\cal N}} \def\cO{{\cal O}}
\def\cP{{\cal P}} \def\cQ{{\cal Q}}
\def\cR{{\cal R}} \def\cV{{\cal V}}\def\cW{{\cal W}}
%
%
%
\def\crr{\crcr\noalign{\vskip {8.3333pt}}}
\def\tilde{\widetilde}
\def\bar{\overline}
\def\us#1{\underline{#1}}
\let\shat=\hat
\def\hat{\widehat}
\def\hyp{\vrule height 2.3pt width 2.5pt depth -1.5pt}
\def\square{\mbox{.08}{.08}}
\def\Coeff#1#2{{#1\over #2}}
\def\Coe#1.#2.{{#1\over #2}}
\def\coeff#1#2{\relax{\textstyle {#1 \over #2}}\displaystyle}
\def\coe#1.#2.{\relax{\textstyle {#1 \over #2}}\displaystyle}
\def\half{{1 \over 2}}
\def\shalf{\relax{\textstyle {1 \over 2}}\displaystyle}
\def\dag#1{#1\!\!\!/\,\,\,}
\def\to{\rightarrow}
\def\notin{\hbox{{$\in$}\kern-.51em\hbox{/}}}
\def\shdot{\!\cdot\!}
\def\ket#1{\,\big|\,#1\,\big>\,}
\def\bra#1{\,\big<\,#1\,\big|\,}
\def\equaltop#1{\mathrel{\mathop=^{#1}}}
\def\Trbel#1{\mathop{{\rm Tr}}_{#1}}
\def\inserteq#1{\noalign{\vskip-.2truecm\hbox{#1\hfil}
\vskip-.2cm}}
\def\attac#1{\Bigl\vert
{\phantom{X}\atop{{\rm\scriptstyle #1}}\phantom{X}}}
\def\exx#1{e^{{\displaystyle #1}}}
\def\del{\partial}
\def\delbar{\bar\partial}
\def\nex#1{$N\!=\!#1$}
\def\dex#1{$d\!=\!#1$}
\def\cex#1{$c\!=\!#1$}
\def\eg{{\it e.g.}} \def\ie{{\it i.e.}}
%
\def\cS{{\cal K}}
\def\IE{\relax{{\rm I\kern-.18em E}}}
\def\cE{{\cal E}}
\def\rt{{\cR^{(3)}}}
\def\IGam{\relax{{\rm I}\kern-.18em \Gamma}}
\def\IGa{\IA}
\def\LG{Lan\-dau-Ginz\-burg\ }
\def\cV{{\cal V}}
\def\Rt{{\cal R}^{(3)}}
\def\wabc{W_{abc}}
\def\WABC{W_{\a\b\c}}
\def\W{{\cal W}}
\def\tft#1{\langle\langle\,#1\,\rangle\rangle}
\def\IA{\relax{\hbox{{\rm A}\kern-.82em {\rm A}}}}
\let\picfuc=\fp
\def\hata{{\shat\a}}
\def\hatb{{\shat\b}}
\def\hatA{{\shat A}}
\def\hatB{{\shat B}}
\def\bv{{\bf V}}
\def\spg{special geometry}
\def\sc{SCFT}
\def\leel{low energy effective Lagrangian}
\def\pf{Picard--Fuchs}
\def\pfS{Picard--Fuchs system}
\def\el{effective Lagrangian}
\def\Fb{\overline{F}}
\def\nablab{\overline{\nabla}}
\def\Ub{\overline{U}}
\def\Db{\overline{D}}
\def\zb{\overline{z}}
\def\eb{\overline{e}}
\def\fb{\overline{f}}
\def\tb{\overline{t}}
\def\Xb{\overline{X}}
\def\Vb{\overline{V}}
\def\Cb{\overline{C}}
\def\Sb{\overline{S}}
\def\delb{\overline{\del}}
\def\Gammab{\overline{\Gamma}}
\def\Ab{\overline{A}}
\def\Anh{A^{\rm nh}}
\def\alphab{\bar{\alpha}}
\def\cy{Calabi--Yau}
\def\cabg{C_{\alpha\beta\gamma}}
\def\B{\Sigma}
\def\Bh{\hat \Sigma}
\def\Kh{\hat{K}}
\def\Knh{{\cal K}}
\def\A{\Lambda}
\def\Ah{\hat \Lambda}
\def\R{\hat{R}}
\def\V{{V}}
\def\T{T}
\def\Gammah{\hat{\Gamma}}
\def\twot{$(2,2)$}
\def\K{K\"ahler}
\def\rat{({\theta_2 \over \theta_1})}
\def\lv{{\bf \omega}}
\def\w{w}
\def\CP{C\!P}
\def\o#1#2{{{#1}\over{#2}}}
\newcommand{\be}{\begin{equation}}
\newcommand{\ee}{\end{equation}}
\newcommand{\ba}{\begin{eqnarray}}
\newcommand{\ea}{\end{eqnarray}}
\newtheorem{definizione}{Definition}[section]
\newcommand{\bd}{\begin{definizione}}
\newcommand{\ed}{\end{definizione}}
\newtheorem{teorema}{Theorem}[section]
\newcommand{\bth}{\begin{teorema}}
\newcommand{\eth}{\end{teorema}}
\newtheorem{lemma}{Lemma}[section]
\newcommand{\blem}{\begin{lemma}}
\newcommand{\elem}{\end{lemma}}
\newcommand{\brr}{\begin{array}}
\newcommand{\err}{\end{array}}
\newcommand{\nn}{\nonumber}
\newtheorem{corollario}{Corollary}[section]
\newcommand{\bcorol}{\begin{corollario}}
\newcommand{\ecorol}{\end{corollario}}
\def\twomat#1#2#3#4{\left(\begin{array}{cc}
 {#1}&{#2}\\ {#3}&{#4}\\
\end{array}
\right)}
\def\twovec#1#2{\left(\begin{array}{c}
{#1}\\ {#2}\\
\end{array}
\right)}
\begin{titlepage}
\hskip 12cm
\vbox{\hbox{CERN-TH/97-95}\hbox{hep-th/9705024} \hbox{May, 1997}}
\vfill
\begin{center}
{\LARGE {Five Dimensional U-Duality, Black-Hole Entropy and
Topological Invariants}}\\
\vskip 1.5cm
{  {\bf Laura Andrianopoli$^1$,
Riccardo D'Auria$^2$ and
Sergio Ferrara$^3$ }} \\
\vskip 0.5cm
{\small
$^1$ Dipartimento di Fisica, Universit\'a di Genova, via Dodecaneso 33,
I-16146 Genova\\
and Istituto Nazionale di Fisica Nucleare (INFN) - Sezione di Torino, Italy\\
\vspace{6pt}
$^2$ Dipartimento di Fisica, Politecnico di Torino,\\
 Corso Duca degli Abruzzi 24, I-10129 Torino\\
and Istituto Nazionale di Fisica Nucleare (INFN) - Sezione di Torino, Italy\\
\vspace{6pt}
$^3$ CERN, Theoretical Division, CH 1211 Geneva, 23 \\
Department of Physics and Astronomy, Rutgers University,
Piscataway, NJ 08855-0849 \\
The Rockfeller University, 1230 York Avenue, N.Y. 10021
}
\end{center}
\vfill
\begin{center} {\bf Abstract}
\end{center}
{
\small
We find the topological entropy formula for all $N>2$ extended supergravities
 in five dimensions and the fixed scalar condition for the black-hole
 ``potential
energy'' which extremizes the BPS mass.
We comment upon the interpretation of these results in a string and
M-theory setting.
}
\vspace{2mm} \vfill \hrule width 3.cm
{\footnotesize
\noindent
$^*$ Work supported in part by EEC under TMR contract ERBFMRX-CT96-0045
 (LNF Frascati,
Politecnico di Torino and Univ. Genova) and by DOE grant
DE-FGO3-91ER40662}
\end{titlepage}
\section{Introduction}
Recently, considerable attention has been given to macroscopic
\cite{black}-\cite{kall}
and microscopic \cite{dbr}, \cite{micros} properties  of
black-holes arising in the low energy approximation of string theory and
M-theory.

In particular, the use of U-duality and the determination of some
black-hole quantum attributes,
such as their mass, charge and entropy, has dramatically come to
play an important role.

The entropy of BPS-extremal black-holes, both in four and five
dimensions, has been computed,
in the macroscopic theory, using a fairly general principle, namely
the extremization of
the BPS mass in moduli space.
This extremum also defines  the so called ``fixed-scalars''
whose microscopic properties
have been further studied in recent time \cite{kr}.

In a recent paper \cite{uns} we have considered theories with more than
two supersymmetries
in $D=4$ and extended the analysis of fixed scalars to the general case.

In particular we have found that, although the scalars may generally be
fixed, for arbitrary
magnetic and electric charge configurations only a subset of them gets a
mass in the
black-hole background.

Moreover, topological invariants have been introduced which give a definition
of the entropy
without need of extremization of the BPS mass.

It is the aim of the present paper to extend this analysis in the $D=5$
dimensional case,
for theories with $N>2$ supersymmetries.

Theories with $N=2$ at $D=5$ have been considered earlier \cite{feka1}
and fixed scalars recently analyzed in great detail \cite{kal}.

A technical important difference in this case is that although matter
charges vanish for fixed scalars,
preserving one  supersymmetry requires that the eigenvalues of the central
 charges which are not the BPS mass do not generally vanish at the horizon,
 but are all equal
and fixed in terms of the entropy.

This paper is organized as follows:
in section 2 we discuss the geometry of $N>2$ five dimensional theories,
their U-duality symmetry, the black-hole potential and the differential relations
obeyed by the central charges.
The fixed scalar condition in $D=5$ is then extablished by looking at the extrema
of the black-hole potential of the geodesic action.

In section 3 we derive the topological invariants for $N=4,6,8$ theories
in terms of which the Bekensein-Hawking entropy can be written as
a manifestly
duality invariant expression in terms of the quantized charges.

In section 4 we comment on the relation of these results with string
and M-theories.

\section{The attractor point condition}
Following the same procedure used in $D=4$ \cite{uns}, we give in this section
the expression for the entropy for $N=4, 6, 8$ black-holes in five dimensional
supergravity.
We will exhibit the relation between central and matter charges that
follows from the ``fixed scalar'' condition of ``minimal mass''.

The five dimensional case exhibits analogies and differencies with
respect to the
four dimensional one.

Exactly like in the four dimensional theories,
the entropy is given, through the Bekenstein-Hawking relation,
by an invariant of the U-duality group over the entire moduli
space and its value is given in terms of the moduli-dependent
  scalar potential of the geodesic action\cite{bmgk} at the attractor point
  \cite{fks}, \cite{feka1},
  \cite{feka2}:
\begin{equation}
  \label{entropy}
  S= {A \over 4}={\pi ^2 \over 12} M^{3/2}_{extr}={\pi ^2 \over 12}
  \left[{\sqrt{3} \over 2}
  V(\phi _{fix}, g)\right]^{3/4}
\end{equation}
where we have used the relation $ V_{extr} = {4 \over 3} M^2_{extr}$
which is valid for any $N$ in $D=5$, as we will show in the following.

Note that, while in the four dimensional case the U-invariant is quartic
in the charges, in $D=5$
it turns out to be cubic.

Furthermore, in five dimensions the automorphism group under which the
central charges transform
is $USp(N)$ instead of $SU(N) \times U(1)$ as in the four dimensional
theories \cite{noi}, \cite{cj}.

As it is apparent from the dilatino susy transformation law, this implies that,
 at the minimum of the ADM mass, the central charges
 different from the maximal one do not vanish, contrary to what happens in
 $D=4$.

The vanishing of charges in $D=4$ at the attractor point was also recovered
in ref \cite{uns},
 in an alternative way, by looking
at the extremum of the scalar ``potential'' $V(\phi,p,q)$ of the geodesic
action \cite{bmgk}, \cite{fegika},
where the following equation
holds:
\begin{equation}
  \label{zz}
  P^{ABCD} Z_{AB} Z_{CD} =0 ;\quad Z_I=0
\end{equation}
$P_{ABCD}$ being the vielbein of the scalar manifold, completely antisymmetric
in its $SU(N)$ indices.
It is easy to see that in the normal frame these equations imply:
\begin{eqnarray}
  \label{norm}
  M_{ADM}\vert _{fix} & \equiv & \vert Z_{1} \vert \neq 0 \\
 \vert Z_{i} \vert & =& 0 \qquad (i=2, \cdots ,N/2 )
\end{eqnarray}

In $D=5$  the same kind of procedure  can be applied but, due to the
traceless
condition of the antisymmetric symplectic representations of the vielbein
$P_{ABCD}$
and of the central charges $Z_{AB}$,the  analogous equations:
\begin{equation}
  \label{zz5}
  P^{ABCD} Z_{AB} Z_{CD} =0 ;\quad Z_I=0
\end{equation}
 do not imply anymore the
vanishing of the central charges different from the mass.
 Here $A,B,\cdots $  are $USp(N)$  indices
and  the antisymmetric matrix $Z_{AB}$ satisfies the reality condition:
\begin{equation}
{\bar Z}^{AB} = \IC^{AC}\IC^{BD}Z_{CD}
\end{equation}
$\IC^{AB}$ being the symplectic invariant antisymmetric matrix
satisfying $\IC = -\IC^T$, $ \IC^2 = - \bfone$.

Let us now consider more explicitly the various five dimensional theories.

In $N=4$ matter coupled supergravity \cite{noi}, the scalar manifold is given
by
 $G/H={O(5,n)\over O(5) \times O(n) } \times O(1,1)$.

The black-hole potential is given by:
\begin{equation}
 V(\phi,q) = {1 \over 2} Z_{AB} \bar Z^{AB} + 2X^2 + Z_I Z^I =
 q_\Lambda (\cN ^{-1})^{\Lambda\Sigma} q_\Sigma
\end{equation}
where $X$ is the central charge associated to the singlet photon of
the $N=4$ theory,  $q_\Lambda \equiv \int_{S_3} \cN_{\Lambda\Sigma}
\,^\star F^\Sigma$ and $\cN _{\Lambda\Sigma}$
 are the electric charges
and vector kinetic matrix respectively.

The central charge $Z_{AB}$ can be decomposed in its $\IC$-traceless
 part $\buildrel \circ\over Z_{AB}$ and trace part $X$ according to:
\begin{equation}
  \label{centr4}
  Z_{AB} = { \buildrel \circ\over Z}_{AB} - \IC_{AB} X
\end{equation}
This decomposition corresponds to the combination of the five graviphotons
and the singlet photon
appearing in the gravitino transformation law \cite{noi}.
The matter charges $Z_I$ are instead in the vector representation of $O(n)$.

Note that in the dilatino transformation law a different
 combination of the five graviphotons and the singlet photon appears
corresponding to the integrated charge:
\begin{equation}
  \label{zchi}
 Z^{(\chi)}_{AB}\equiv {1 \over 2}( Z_{AB} + 3 \, \IC_{AB}X )=
 {1 \over 2}({ \buildrel \circ\over Z}_{AB} + 2
 \, \IC_{AB}X )
\end{equation}

The differential relations satisfied by the central and matter charges are:
\begin{eqnarray}
  \label{difrel4}
  \nabla Z_{AB} = P_{IAB} Z^I - 2 Z^{(\chi)}_{AB}d\sigma  & \to &
  \nabla {\buildrel \circ\over Z}_{AB} = P_{IAB} Z^I -
  {\buildrel \circ\over Z}_{AB}d\sigma \\
\nabla X & =& 2 X d\sigma \\
\nabla Z_I &=& {1\over 4}( Z_{AB}\bar P^{AB}_I + \bar Z^{AB}P_{AB I})
- Z_I d \sigma
\end{eqnarray}

To minimize the potential, it is convenient to go to the normal frame
where ${\buildrel \circ \over Z}_{AB}$ has proper values $e_1, e_2 = - e_1$.
In this frame, the potential and the differential relations become:
\begin{eqnarray}
V(\phi, q) &=& e_1^2 + e_2^2 + 4 X^2 \label{vnor4} \\
\nabla e_1 &=& - e_1 d \sigma  + P_{I} Z^I \\
 \nabla X & =& 2 X d\sigma
\end{eqnarray}
where $P_I \equiv P_{I 12}$ is the only independent component of the
traceless vielbein one-form $P_{I AB}$ in the normal frame.
We then get immediately that,
in the normal frame, the fixed scalar condition
${\partial V \over \partial \phi}=0$ requires:
\begin{eqnarray}
  Z_I&=& 0 \\
 {  e}_{1} &=&- { e}_{2}=  -2 X
  \end{eqnarray}
where ${ e}_{i}$ ($i=1,2$) are the proper  values of
${ \buildrel \circ\over Z}_{AB}$.
It follows that $M_{extr}= \vert Z_{12\ extr} \vert =  {3 \over 2} e_1$
so that
\begin{equation}
V_{extr}= 3 e_1^2 = {4 \over 3} M_{extr} ^2
\end{equation}

In the $N=6$ theory, the scalar manifold is $G/H = SU^\star (6)/Sp(6)$
\cite{noi}.
The central charge $Z_{AB}$ can again be  decomposed in a $\IC$-traceless
 part $\buildrel \circ\over Z_{AB}$ and a trace part $Z$ according to:
\begin{equation}
  \label{centr6}
  Z_{AB} = { \buildrel \circ\over Z}_{AB} + {1 \over 3} \, \IC_{AB} X
\end{equation}

The traceless and trace parts satisfy the differential relations:
\begin{eqnarray}
  \label{difrel6}
  \nabla { \buildrel \circ\over Z}_{AB} &=&
  { \buildrel \circ\over Z}_{C[A}P_{B]D} \, \IC^{CD}
  + {1\over 6}\, \IC_{AB}{ \buildrel \circ\over Z}_{LM}P^{LM}
  +{2 \over 3} XP_{AB}\\
\nabla X &=&{1\over 4} { \buildrel \circ\over Z}_{AB}P^{AB}
\end{eqnarray}
where $P_{AB}$ is the $\IC$ traceless vielbein of $G/H$.

The geodesic potential has the form:
\begin{equation}
  \label{pot6}
  V(\phi,q)= {1\over 2} Z_{AB} Z^{AB} +  X^2= q_{\Lambda\Sigma}
  (\cN ^{-1})^{\Lambda\Sigma , \Gamma\Delta}
  q_{\Gamma\Delta}
\end{equation}
 where $ \Lambda,\Sigma , \cdots = 1, \cdots ,6$ are indices in
 the fundamental representation
 of $SU^\star (6)$, the couple of indices $\Lambda\Sigma$ in the
 elctric charges
 $q_{\Lambda\Sigma}$ are antisymmetric and $ (\cN )_{\Lambda\Sigma
 , \Gamma\Delta} $ is
 the kinetic matrix  of the vector field-strengths
 $F^{\Lambda\Sigma}$
To perform the minimization of the potential we proceed as in the
$N=4$ case going to the normal frame where ${ \buildrel \circ\over Z}_{AB}$
has proper values  $e_1, e_2, e_3=-e_1-e_2$.
The potential becomes:
  \begin{equation}
  \label{potnor6}
  V(\phi,q)=  e_1^2 + e_2^2 + (e_1 + e_2 )^2 + {4 \over 3}X^2
\end{equation}
   Moreover, the differential relations (\ref{difrel6}) take the
   form:
 \begin{eqnarray}
\nabla e_1 &=& {1\over 3} (- e_1+ e_2 + 2 X)P_1 + {1\over 3}(e_1 + 2
e_2) P_2 \label{difnor61}\\
 \nabla e_2 &=&   {1\over 3}(2e_1 +
e_2) P_1  + {1 \over 3} ( e_1 - e_2 + 2 X)P_2 \label{difnor62}\\
 \nabla X &=& {1 \over 2} ( 2e_1 + e_2)P_1 + {1 \over 2} (e_1 + 2
e_2) P_2
\label{difnor63}
\end{eqnarray}
 where $P_1,P_2,P_3=-P_1 -
P_2$ are the proper values of the vielbein one-form
$P_{AB}$ in the normal frame.

Imposing the attractor-point constraint ${\partial V \over \partial \phi} =0$
on the potential we get the following
relations among the charges at the extremum:
\begin{equation}
 { e}_{2} =   { e}_{3} =  - {1\over 2} {  e}_{1}=  - {4 \over 3}X  \,
 ; \quad V_{extr} = {27 \over 16} e_1^2 = {4 \over 3} M_{extr}^2 .
 \label{min6}
  \end{equation}

Note that, using eqs. ( \ref{difnor61})-(\ref{difnor63}), (\ref{min6}),
the mass $e_1 + {1 \over 3} X $
satisfies ${ \partial \over \partial \phi^i} (e_1 + {1 \over 3} X  )
=0$ at the extremum,
with value $M_{extr} = {9 \over 8} e_1$.
\vskip 0.5cm

In the $N=8$ supergravity the scalar manifold is $G/H = E_{6(-6)}/Sp(8)$
and the central charges sit in the  twice antisymmetric, $\IC$-traceless,
representation of $USp(8)$ \cite{noi}.

The scalar ``potential'' of the geodesic action is given by:
\begin{equation}
  \label{pot8}
  V(\phi , g) = {1 \over 2} Z_{AB} \bar Z^{AB} = q_{\Lambda\Sigma}
  (\cN^{-1}) ^{\Lambda\Sigma, \Gamma\Delta}(\phi ) q_{\Gamma\Delta}
\end{equation}
where $q_{\Lambda\Sigma} \equiv \int\cN_{\Lambda\Sigma ,
\Gamma \Delta}F^{\Gamma\Delta}$
are the electric charges
 and $\cN _{\Lambda\Sigma, \Gamma\Delta}$ the vector kinetic matrix.

The extremum of $V$ can be found by using the differential relation
\cite{noi}:
\begin{equation}
  \label{difrel}
  \nabla Z_{AB} ={1 \over 2} P_{ABCD} \bar Z^{CD}
\end{equation}
where $P_{ABCD}$ is the four-fold antisummetric vielbein one-form of
$G/H$.
One obtains:
\begin{equation}
 P^{ABCD} Z_{AB} Z_{CD} =0
\end{equation}

To find the values of the charges at the extremum we use the
traceless conditions
$\IC^{AB} P_{ABCD} =0$, $\IC^{AB} Z_{AB} =0$.
In the normal frame, the proper values of $Z_{AB}$ are
$e_1, e_2, e_3, e_4=-e_1 -e_2- e_3$  and we take, as
independent components of the vielbein,
$P_1 = P_{1234} = P_{5678}$, $P_2 = P_{1256}  = P_{3478}$ ($P_{3456}=
P_{1278} = - P_1 - P_2$).
In this way, the covariant derivatives of the charges become:
\begin{eqnarray}
\label{difnor8}
\nabla e_1 = (e_1 + 2 e_2 + e_3) P_1 + (e_1 + e_2 + 2 e_3) P_2 \\
\nabla e_2 = (e_1 - e_3) P_1 + (-e_1 - e_2 - 2 e_3) P_2 \\
\nabla e_3 = (-e_1 - 2 e_2 - e_3) P_1 + (e_1 - e_2) P_2 \\
\end{eqnarray}

Using these relations, the extremum condition of $V$ implies:
\begin{equation}
  \label{extrz}
  e_2 = e_3 = e_4 = -{1\over 3} e_1 \, ; \quad  V_{extr} =
  {4 \over 3} e_1^2 = {4 \over 3} M_{extr}^2
\end{equation}

\section{Topological invariants}
In this section we determine the U-invariant expression in terms of
which the entropy can be evaluated over the entire moduli space.

Our procedure is the same as in ref. \cite{uns}, namely we compute
cubic H-invariants
and determine the appropriate linear combination of them which turns
out to be U-invariant.

The invariant expression of the entropy for $N=4$ and $N=8$ at $D=5$
in terms of the quantized charges was given in \cite{feka2}.

Let us begin with the $N=4$ theory, controlled by the coset
${O(5,n)\over O(5)\times O(n)} \times O(1,1)$.

In this case there are three possible cubic H-invariants, namely:
 \begin{eqnarray}
I_1 &=& {1\over 2} {\buildrel \circ \over Z}_{AB}  {\buildrel \circ
\over {\bar Z}}^{AB}X   \\
 I_2 &=& Z_I Z_I X \\
I_3 &=& X^3
\end{eqnarray}
In order to determine the $U \equiv G= O(5,n) \times O(1,1)$-invariant,
we set $I=I_1 + \alpha I_2 + \beta I_3$ and
  using the differential  relations  (\ref{difrel4}) one easily
finds that $\nabla I =0$ implies $\alpha =1$, $\beta =0$.

Therefore:
\begin{equation}
I= I_1 - I_2 = \left({1\over 2} \buildrel\circ\over Z_{AB}
{\buildrel\circ\over {\bar Z}}^{AB} - Z_I Z_I \right)X
\end{equation}
 is the cubic ($O(5,n) \times O(1,1)$)-invariant independent
of the moduli and the entropy $S={\pi \over 12} M^{3\over 2}$ is
then given as:

\begin{equation}
S \sim I^{1/2}= \sqrt{({1\over 2} Z_{AB}\bar Z^{AB} -Z_I Z_I)Z }
\label{cubinv4}
\end{equation}

In the $N=6$ theory, where the coset manifold is $SU^*(6) /Sp(6)$,
the possible cubic $Sp(6)$-invariants are:
\begin{eqnarray}
I_1&=& Tr(Z\IC)^3 \\
 I_2&=& Tr(Z\IC)^2X \\
 I_3&=& X^3
\end{eqnarray}

Setting as before:
\begin{equation}
I= I_1 + \alpha I_2 + \beta I_3
\end{equation}
the covariant derivative $ \nabla I$ is computed using the differential
relations (\ref{difrel6}) and the parametrere
$\alpha$ and $\beta$ are then determined by imposing $\nabla I =0$.
Actually, the best way to perform the computation is to go to the
normal frame.
Using the differential relations (\ref{difnor61})-(\ref{difnor63})
and the expression for the invariants in the normal frame:
\begin{eqnarray}
Tr(Z\IC)^3 &=& - 6 (e_1^2 e_2 + e_1 e_2 ^2)\\
Tr(Z\IC)^2 &=& 4 (e_1^2+ e_2^2 + e_1 e_2 )
\end{eqnarray}
  the vanishing of $\nabla I$ fixes the coefficients
$\alpha$ and $\beta$.
The final result is:
\begin{equation}
I=-{1\over 6}Tr( Z\IC)^3 - {1\over 6} Tr( Z\IC)^2 X + {8 \over 27} X^3
\label{cubinv6}
\end{equation}
and the entropy is:
\begin{equation}
S \sim \sqrt{ -{1\over 6}Tr( Z\IC)^3 -  {1\over 6}Tr( Z\IC)^2 X +
{8 \over 27} X^3}
\label{cubinv8}
\end{equation}

Finally, for the $N=8$ theory, described by the coset
$E_{6(6)}/Sp(8)$,  it is well known that there is  a unique $E_{6}$
cubic invariant, namely:
\begin{equation}
I=Tr(Z\IC )^3= Z_{A}^{\ B} Z_{B}^{\ C} Z_{C}^{\ A}
\end{equation}
where the $Sp(8)$ indices are raised and lowered by the antisymmetric
matrix $\IC_{AB}$.

Curiously, the $E_6$-invariant corresponds to a single cubic
$USp(8)$-invariant.

Again, the invariance of $I$ can be best computed in the normal
frame where the   invariant (\ref{cubinv8})  takes the form:
\begin{equation}
 Tr(Z\IC )^3= -2(e_1^3 + e_2 ^3 + e_3^3 + e_4 ^3) = 6(e_2^2 e_3+ e_2
 e_3 ^2 + e_1^2e_2 + e_1e_2^2+ e_1^2e_3 + e_1e_3^2+ 2 e_1e_2 e_3)
\end{equation}
One finds indeed $\nabla I=0$ and therefore
\begin{equation}
S \sim I^{1/2} = \sqrt{Tr(Z\IC)^3}
\end{equation}

We note that in the normal frame the transformations of the coset
which preserve the normal form of the $Z_{AB}$ matrix belong to
$O(1,1)^2$ both for $N=6$ and $N=8$ theories.

The relevant $O(1,1)^2$ transformations can be read out from eqs.
(\ref{difnor61})-(\ref{difnor63}), (\ref{difnor8}) by replacing
the vielbein $P_1,P_2$
with parameters $\xi_1, \xi_2$ and the $\nabla$ simbol with the
variation simbol $\delta$.

The ensuing transformations correspond to the following commuting
matrices $A$, $B$ which are proper, non compact, Cartan elements of the coset
algebra $\IK$:
\begin{equation}
N=6 : \quad  A={1 \over 3}\pmatrix{ -1 & 1 & 2 \cr 2 & 1 & 0 \cr 3
& {3 \over 2} & 0 \cr} ; \quad B={1 \over 3} \pmatrix{1 & 2 & 0 \cr 1 & -1 &
2 \cr {3 \over 2} & 3 & 0 \cr}
\end{equation}
\begin{equation}
N=8 : \quad  A={1 \over 2}\pmatrix{ 1 & 2 & 1 \cr 1 & 0 & -1 \cr  -1
& -2 & -1\cr} ; \quad B= {1 \over 2}\pmatrix{1 & 1 & 2 \cr -1 & -1 &
-2 \cr 1 & -1 & 0 \cr}
\end{equation}
In both cases the matrices $A$, $B$ satisfy $A^3 =A$, $B^3=B$.

 \section{Relation to string and M-theory}
 Theories of the type described in the present paper have been
 actually considered for the microscopic determination of the mass
 and entropy of black-holes arising in string theory and M-theory.

$N=2$ black-holes at $D=5$ \cite{cfgk} arise in M-theory compactified
on a Calabi-Yau threefold or heterotic string on $K_3 \times S_1$
\cite{cad}-\cite{BBS}.
In this case the moduli space geometry is ``very special geometry''
\cite{gst}, \cite{dwvp} which does not require (local) homogeneous
spaces.

From $N=8$, by decomposition to $N=2$ at $D=5$ we obtain a gravitino
multiplet with a spin $3/2$ in the 6 of $Sp(6)$, 14 vector multiplets
with associated coset $SU^\star (6) / Sp(6) $ \cite{cj} and 7 hypermultiplets
with associated exceptional coset $F_{4(4)}/Sp(2) \times Sp(6)$.

This means that ${E_{6(6)}\over USp(8)}= {SU^\star (6)\over USp(6)}
+{F_{4(4)}\over USp(2) \times USp(6)}$ in the sense of solvable Lie
algebras \cite{solv}.

This also means that, by truncating $N=8$ at $D=5$ to $N=2$, we may
get a theory with $n_V=14$, $n_H=0$ or a theory with $n_V=0$,
$n_H=7$.

If we compactify M-theory on $T_6 /Z_3$ we get for example $n_V=8$
with $\cM_{V} = {SL(3,\IC) / SU(3)}$ and $n_H=1$ with $\cM_H =
{SU(2,1) \over SU(2) \times U(1)}$.

There is a series of very special manifolds for each $Z_N$ orbifold,
since we get a $N=2$ theory at $D=5$ by compactifying M-theory on any
$Z_N$ orbifold.

These manifolds are given in ref. \cite{gst}.

As in four dimensions, fixed scalars can be analyzed in terms of
$N=2$ multiplets \cite{uns}.
Therefore we may expect that, in $N=8$, 28 scalars remain massless, while
14 get a mass from the geodesic potential.

A similar analysis can be done for lower $N$ theories.

$N=4$ appears in heterotic string on $T_5$ or M-theory on
$K_3 \times T_2$.
In both cases the number of matter multiplets is $n=21$ and the
$O(5,21;\ZZ)$ U-duality group is related to the Narain lattice in
heterotic or to the $K_3$ cohomology structure in Type II.

$N=8$ corresponds to Type II on $T_6$ or M-theory on $T_7$, while
$N=6$ can be obtained in Type II compactified on a asymmetric
orbifold \cite{fkff}.

``Fixed scalars'' and entropy formulae have been obtained using
D-branes techniques, by microscopic calculations, in these theories \cite{kr}.

It would be interesting to compare  microscopic results in particular
theories with the model-independent analysis performed in this paper.

\end{document}